\documentclass[10pt,letterpaper]{article}
\usepackage{ol2}
\usepackage{varioref}
\usepackage{graphicx}
\usepackage[T1]{fontenc}
\usepackage{subfigure}
\usepackage{cite}
\usepackage{amsmath}
\usepackage{amsfonts}
\usepackage{stmaryrd}
\usepackage{url}

\begin{document}

\title{Individual plasmonic helix for probing light chirality}

\author{Mengjia Wang, Roland Salut, Miguel Angel Suarez, Nicolas Martin and Thierry Grosjean*}

\address{FEMTO-ST Institute UMR 6174, Univ. Bourgogne Franche-Comte --CNRS -- Besancon, France}

\email{*thierry.grosjean@univ-fcomte.fr}

\begin{abstract}
We investigate the plasmonic nanohelix as an individual subwavelength element for locally probing light chirality. We show that an hybrid nanoantenna combining a carbon-gold core-shell helix and a plasmonic nanoaperture transmits circularly polarized light with the same handedness as the helix and blocks the other. Such an assymmetric response is spatially localized, spectrally broadband and background-free. Finally, we demonstrate the possibility to engineer an individual plasmonic helix at the apex of a sharp tip typically used in scanning near-field microscopies, thus opening the prospect of moveable local probes for high resolution sensing and mapping of light chirality and chiroptical forces.
\end{abstract}

\maketitle

\newpage

Chiral metamaterials and nanophotonics have recently attracted much interest for their unique ability to enhance and control inherently weak chiral light-matter interaction \cite{hentschel:sciadv17,schaferling:prx12}. Research in these domains is essentially relying upon the ability to produce and control spinning light through the design of specific subwavelength systems. Characterizing light chirality is crucial to assess and improve such designs. To that end, theoretical modeling has been preparing for a long time with the development of rigorous numerical methods which allow full-wave light field simulation around arbitrarily complex nanodevices. From an experimental point-of-view, the chiroptical properties of nanostructures and metamaterials are usually detected via ensemble-averaged measurements with large, at best diffraction-limited illumination \cite{poulikakos:nl18,meinzer:prb12}. 

Any individual chiral molecule or nanostructure is potentially a local probe of light chirality as it interacts differently to right and left spinning optical fields. However, reaching completely detangled information from the different handedness of light seems to be challenging at the scale of an individual subwavelength chiral element. Background problems and modest chiroptical responses may impede the development of individual nanoscale chiral probes without any complex signal post-processing involving bulky optics. 

Aperture nanoantennas have been recently demonstrated to accurately probe light fields at subdiffraction scale \cite{taminiau:nl07,burresi:science09,mivelle:nl14,murphy:ox08,grosjean:ox10,vo:ox12,lindquist:scirep13,xie:nl17}. These subwavelength structures have been integrated at the apex of different kinds of tips of scanning near-field microscopies to be approached and raster-scanned across a sample while picking up detailed and background-free near-field optical information. The local detection of light chirality with individual subwavelength aperture nanoantennas has been hardly addressed.

Strong chiroptical responses have been obtained from helical nanoantennas \cite{kaschke:nanophotonics16}, leading to giant circular dichroism \cite{gibbs:apl13,esposito:acsphot14,esposito:natcom15}, strongly polarization-dependent transmission \cite{gansel:sci09,kosters:acsphot17} and superchiral light \cite{schaferling:acsphot14}. Core-shell nanohelices have been recently proposed to enhance chiroptical effects at visible frequencies by manipulating surface plasmons  \cite{kosters:acsphot17}. Individual resonant core-shell helices have been shown to sustain chiral dipoles associated with resonances \cite{wozniak:ox18,wozniak:optica19}. In a non-resonant operation, the carbon-gold core-shell plasmonic helix has led to the concept of helical traveling-wave nano-antenna (HTN) enabling subwavelength polarization optics and a new regime in polarization control \cite{wang:lsa19}. 

In this paper, we investigate the carbon-gold core-shell helix as a broadband subwavelength probe of light chirality. When coupled to a nanoaperture in an HTN configuration \cite{wang:lsa19}, a left-handed carbon-gold core-shell is shown to transmit light of left circular polarization and blocks the other. A differential transmission larger than 0.96 is obtained for wavelengths ranging from 1.47 to 1.65 $\mu$m.  Finally, we show that individual carbon-gold core-shell helices can be fabricated at the apex of a tip used in scanning near-field microscopy. This opens the prospect of on-tip local probes for measuring light chirality around nanostructures via direct optical detection or chiroptical force sensing \cite{zhao:natnano17}.\\


We fabricated individual carbon-gold core-shell helices to operate at telecommunication wavelengths ($\lambda$ around 1.55 $\mu$m). The carbon helical skeleton is sculpted by focused ion beam induced deposition (FIBID) \cite{esposito:nl16,wang:lsa19}. Metal coating is realized by sputter-depositing a thin layer of gold onto the carbon core with a metal target tilted with an angle of 80$^{\circ}$ from the helix axis and rotated during the deposition time with a constant rotating speed of 2 rev.min$^{-1}$. We combined these two techniques to fabricate plasmonic helices consisting of a 105-nm diameter carbon wire wound up in the form of a four-turn corkscrew-type structure and covered with a 25-nm thick gold layer. The resulting helix has a 505-nm outer diameter and is 1.66-$\mu$m high.  It is positioned on a cylindrical pedestal of 105 nm diameter and 100 nm high carbon whose lateral side is also coated with a 25-nm thick gold layer. 

\begin{figure}[ht!]
\centering
\includegraphics[width=0.7\linewidth]{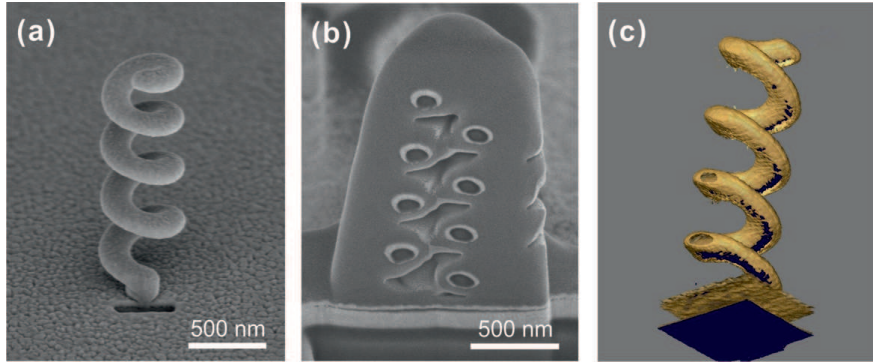}
\caption{(a) Scanning electron micrograph of a gold-coated carbon helix coupled to a rectangle nanoaperture: HTN configuration. (b) Scanning electron micrograph of a cross-section of the helix initially embedded in platinum by FIBID. (c) Reconstructed 3D image of the fabricated gold-coated carbon helix via FIB/SEM tomography of the platinum embedded nanostructure. Yellow: gold; black: carbon.}\label{fig:fab1}
\end{figure}

We considered this plasmonic helix as a part of a non-resonant helical traveling-wave nanoantenna \cite{wang:lsa19}. To this end, the core-shell structure is engineered onto a 100-nm thick gold layer deposited onto a glass substrate and a 370 nm-by-40 nm rectangle nano-aperture is engraved by focused ion beam (FIB) milling in the flat gold layer right at the helix pedestal (Fig. \ref{fig:fab1}(a)). To evaluate the thickness of the gold layer deposited onto the helical carbon core, we perform tomography of the helix by alternating FIB slicing and SEM imaging of the nanostructure. To this end, the helix is embedded in platinum deposited by FIBID. We see in the helix cross-section of Fig. \ref{fig:fab1}(b) that the gold layer (white rings in the image) is not homogeneously distributed all around the carbon wire (dark circular regions in the SEM image). Figure \ref{fig:fab1}(c) shows a 3D reconstruction of the gold-coated carbon helix revealing a bottom side that is not fully covered with gold. The tilt angle of the gold target of 80$^{\circ}$ creates a shadowing effect in the gold coating process which prevents a a uniform gold layer to be deposited all around the helical wire. Non homogeneously-distributed metal coating around the carbon core mat spectrally redshift the helix optical response \cite{wang:lsa19}.

\begin{figure}[ht!]
\centering
\includegraphics[width=0.65\linewidth]{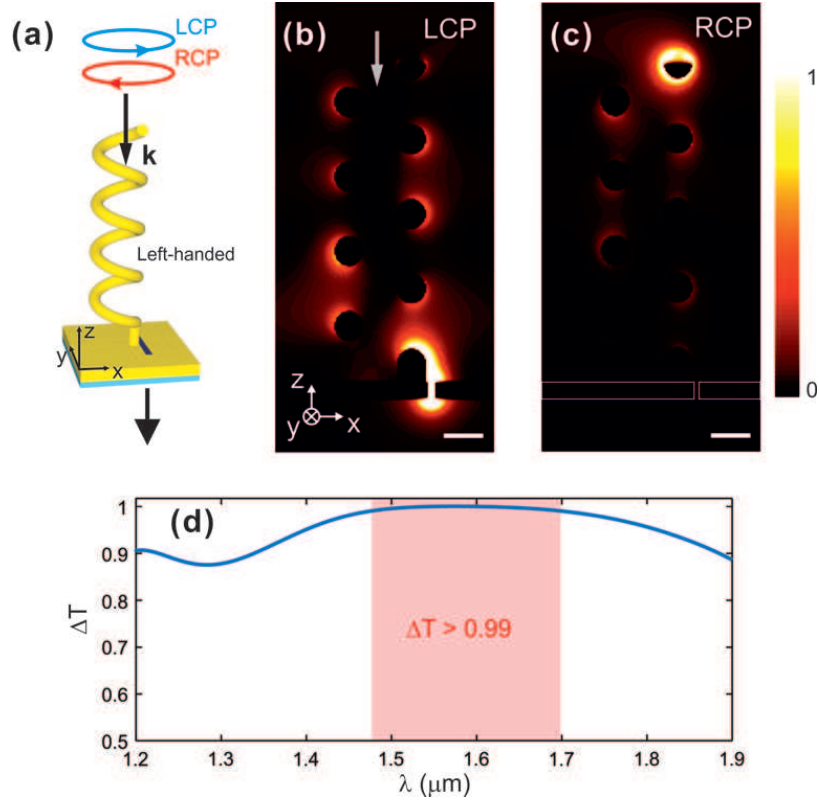}
\caption{(a) Description of the numerical study with the FDTD method. A focused beam is projected onto the helix. The two right and left circular polarizations are considered in two distinct simulations.  (b) and (c) Cross section of the probe-based HTN configuration in the (x0z)-plane with incoming right and left circular polarizations, respectively ($\lambda$=1.57 $\mu$m). Scale bars: 200 nm. (d) Spectrum of the differential transmission $\Delta$T of the probe used in collection mode (simulations in pulsed regime followed by Fourier transform).}\label{fig:simu}
\end{figure}

We performed three-dimensional FDTD simulations using commercial software (Fullwave). The antenna is modeled as a wounded cylindrical carbon wire coated with a 25-nm thick gold layer. The helix is placed next to a rectangle aperture in an extended gold slab lying on a glass substrate. The geometrical parameters of the simulated HTN are those of the structure introduced above. The numerical parameters of the simulations can be found in \cite{wang:lsa19}. The helix is illuminated with a circularly polarized gaussian beam (beam waist: 2.3 $\mu$m) propagating along the helix axis. Because we consider this hybrid nanoantenna as an chiral optical probe working in collection mode, we calculate the transmitted light in the substrate for two incoming circular polarizations of opposite handedness.  An overview is shown in Fig. \ref{fig:simu}(a). The antenna response at $\lambda$=1.57 $\mu$m is reported in Figs. \ref{fig:simu} (b) and (c) for the left and right circular polarizations, respectively. 

The HTN shows an extreme polarization-dependent response as the two right and left circular polarization states induce transmission (Fig. \ref{fig:simu} (b)) and no transmission (Fig. \ref{fig:simu} (c)) of light into the substrate through the nanoaperture.  To quantify such an assymmetric response to the different polarization handedness, we used the differential transmission $\Delta T=(T_{LCP}-T_{RCP})/(T_{LCP}+T_{RCP})$ \cite{kosters:acsphot17}. $T_{LCP}$ and $T_{RCP}$ are the transmission spectra of the HTN used in light collection mode for incoming right and left circular polarizations, respectively. 
Figure \ref{fig:simu} (d) reports on a calculated differential transmission peaking at 1 when $\lambda$=1.57 $\mu$m and remaining larger than 0.99 for wavelengths ranging from 1.48 to 1.7 $\mu$m. Simulations thus predict an extremely assymmetric response of the nanoprobe regarding the right and left circular polarization states, over a broad spectral bandwidth.   

\begin{figure}[ht!]
\centering
\includegraphics[width=0.7\linewidth]{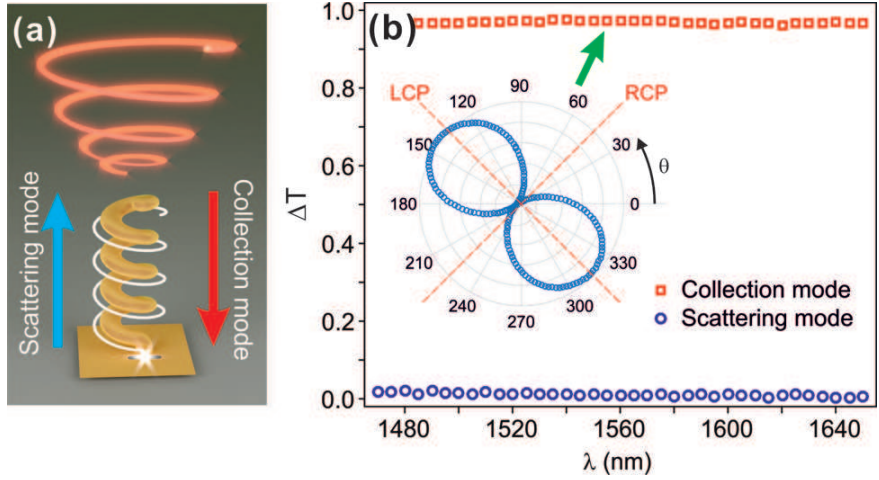}
\caption{(a) Schematics of the two experimentally investigated operations of the HTN, involving light wave propagation in two opposite directions along the helix axis. The collection mode of interest involves helix illumination (red arrow) whereas the scattering mode implies aperture illumination from the backside (blue arrow). (b) Experimental spectra of the differential transmission $\Delta T$ in both the scattering and collection modes of the HTN. Figure inset: transmission through the HTN used in collection mode at $\lambda$=1.55 $\mu$m as a function of the incoming polarization (i.e., the angle $\theta$ between the fast axis of the rotating quarter-wave plate and the direction of the incoming linear polarization).} \label{fig:exp}  
\end{figure}

 
To experimentally verify the polarization-dependent properties of the HTN, we characterized the transmission of the nanoantenna both in the scattering and collection modes represented in Fig. \ref{fig:exp} (a). To this end, linearly polarized light from a tunable laser (Yenista) passes through a rotating quarter-wave plate (QWP; AHWP05M-1600, Thorlabs) before being focused with a (25X, 0.4) microscope objective onto the HTN. The quarter-wave plate is mounted onto a motorized stage (PRM1Z8, Thorlabs) to be accurately rotated with respect to the incident linear polarization direction. The light transmitted through the nanoantenna is detected by imaging the plasmonic structure with an (50X, 0.65) infrared objective from Olympus coupled to an infrared camera (GoldEye model G-033, Allied Vision Technologies GmbH). In the scattering and collection modes of the HTN, the rectangle nano-aperture and the helix are selectively illuminated, respectively. In Fig. \ref{fig:exp} (b) the spectra are obtained by studying the nanoantenna transmission as a function of the wavelength of the incoming light. At each wavelength, the transmitted intensity is measured for two orthogonal orientations of the quarter-wave plate leading to circular right and left polarizations (fast axis of the waveplate at $\pm$45$^{\circ}$ relative to the incident polarization direction). Then, the differential transmission $\Delta T$ is deduced from these measurements. 

We see that $\Delta T$ is almost equal to zero in the scattering mode of the HTN. This can be easily explained by the dipolar nature of the rectangle nano-aperture which imposes a linearly polarized end-firing of the plasmonic helix regardless of the incoming polarization. The nano-aperture acts as a nanoscale linear polarization filter. In collection mode, $\Delta$T remains larger than 0.96 over the whole spectral bandwidth of the laser. This result agrees well with the numerical predictions in Fig. \ref{fig:simu}. It confirms that the HTN develops a strongly assymmetric response with regards to polarization handedness when used as a local probe of light. We also analyze the light collected by the HTN while rotating the quarter-wave plate by one turn. The typical two-lobe pattern of the resulting diagram evidences that in its light collection mode, the HTN blocks the optical waves whose handedness is opposite to the helix, regardless the incoming polarization.


\begin{figure}[ht!]
\centering
\includegraphics[width=0.7\linewidth]{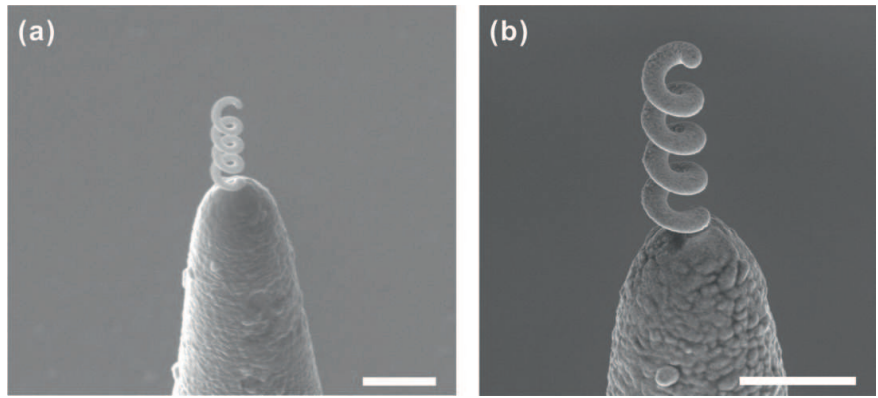}
\caption{Gold-coated carbon helix at the apex of a tip used in scanning near-field microscopy.  (a) Large view of the helix carbon skeleton sculpted by FIBID at the very tip. (b) Finalized on-tip gold-coated carbon helix. Scale bars: 1 $\mu$m.}\label{fig:tip}  
\end{figure}

Local probing of light chirality requires moveable nanoprobes. We finally investigate the possibility to integrate an individual gold-coated carbon helix at the apex of the sharp tips used in scanning near-field microscopies. To this end, we considered a typical model of a sharp dielectric near-field tip. We metal-coated the tip with a thin layer of aluminum to ensure charge removal during FIBID process. In the case of tips with nanoscale apex such as those used in atomic force microscopy, a flat of 200-300 nm width can be engineered by FIB at the very tip prior to FIBID to ensure a reliable fabrication process of the nanostructures. We performed on-tip fabrication of the above introduced carbon-gold core-shell helix with the same parameters of the FIBID and metal coating as in the previous on-chip fabrication \cite{wang:lsa19}. Figure \ref{fig:tip} shows SEM images of a resulting tip-integrated structure before (Figs. \ref{fig:tip}(a)) and after (Figs. \ref{fig:tip}(b)) metal coating.  Changing the helix holder does not affect much the fabrication process. FIBID keeps its high level of accuracy in the definition of the helix geometry and metal coating still leads to smooth gold layers compatible with plasmonics. These images demonstrate the on-tip integration of an individual plasmonic helix aimed at probing light chirality.\\


To conclude, we investigate the core-shell nanohelix as an individual subwavelength element for locally probing light chirality. By coupling an helix to a nano-aperture engraved in an opaque metallic layer, we obtained background-free light collection on the subwavelength scale. Probing light chirality can thus be realized without any signal post-processing involving bulky optical components. Such a local probe can be developed directly onto a photodetector or at the apex of a fiber tip used in scanning near-field optical microscopy. We study here the gold-coated carbon helix as a non-resonant "`travelling-wave"' nano-antenna \cite{balanis:book,kraus:book,wang:lsa19}. In a reciprocal approach of Schaferling's et al. \cite{schaferling:acsphot14}, tuning the helix to the resonant mode \cite{balanis:book,kraus:book,wozniak:ox18,wozniak:optica19} by downscaling the structure would lead to chiral dipoles capable to locally characterize superchiral optical fields and effects.  Coupling these resonant helical structures to coaxial nanoapertures may enable direct background-free local probing of light super-chirality. Coaxial nanoapertures sustain the radially-polarized TEM plasmon mode \cite{banzer:ox10,baida:apb07} which can be coupled to the helix chiral dipole. We finally show the possibility to fabricate a single plasmonic helix at the apex of tips used in near-field microscopies. Beyond direct measurement of light chirality, the integration of a plasmonic helix at the end of an AFM tip should enable the local measurement of enantioselective optical forces \cite{zhao:natnano17}. Plasmonic helices may thus pave the way for a new generation of nanoprobes for locally measuring light chirality and chiroptical effects as well as opto-mechanic, opto-acoustic and magneto-optic phenomena. 

\section*{Funding}
Region Bourgogne Franche-Comte; EIPHI Graduate School (ANR-17-EURE-0002); Agence Nationale de la Recherche: ANR-18-CE42-0016


\begin{thebibliography}{10}

\bibitem{hentschel:sciadv17}
Mario Hentschel, Martin Sch{\"a}ferling, Xiaoyang Duan, Harald Giessen, and
  Na~Liu.
\newblock Chiral plasmonics.
\newblock {\em Sci. Advances}, 3(5):e1602735, 2017.

\bibitem{schaferling:prx12}
Martin Sch{\"a}ferling, Daniel Dregely, Mario Hentschel, and Harald Giessen.
\newblock Tailoring enhanced optical chirality: design principles for chiral
  plasmonic nanostructures.
\newblock {\em Phys. Rev. X}, 2(3):031010, 2012.

\bibitem{poulikakos:nl18}
Lisa~V Poulikakos, Prachi Thureja, Alexia Stollmann, Eva De~Leo, and David~J
  Norris.
\newblock Chiral light design and detection inspired by optical antenna theory.
\newblock {\em Nano Lett.}, 18(8):4633--4640, 2018.

\bibitem{meinzer:prb12}
Nina Meinzer, Euan Hendry, and William~L Barnes.
\newblock Probing the chiral nature of electromagnetic fields surrounding
  plasmonic nanostructures.
\newblock {\em Phys. Rev. B}, 88(4):041407, 2013.

\bibitem{taminiau:nl07}
T.~Taminiau, R.~Moerland, F.~Segerink, L.~Kuipers, and N.~Van Hulst.
\newblock $\lambda$/4 resonance of an optical monopole antenna probes by single
  molecule fluorescence.
\newblock {\em Nano Lett.}, 7:28, 2007.

\bibitem{burresi:science09}
M.~Burresi, D.~van Oosten, T.~Kampfrath, H.~Schoenmaker, R.~Heideman,
  A.~Leinse, and L.~Kuipers.
\newblock Probing the magnetic field of light at optical frequencies.
\newblock {\em Science}, 326:550--553, 2009.

\bibitem{mivelle:nl14}
Mathieu Mivelle, Thomas~S van Zanten, and Maria~F Garcia-Parajo.
\newblock Hybrid photonic antennas for subnanometer multicolor localization and
  nanoimaging of single molecules.
\newblock {\em Nano Lett.}, 14(8):4895--4900, 2014.

\bibitem{murphy:ox08}
N.~Murphy-DuBay, L.~Wang, E.~C. Kinzel, S.~M.~V. Uppuluri, and X.~Xu.
\newblock Nanopatterning using nsom probes integrated with high transmission
  nanoscale bowtie aperture.
\newblock {\em Opt. Express}, 16(4):2584--2589, 2008.

\bibitem{grosjean:ox10}
T.~Grosjean, I.~A. Ibrahim, M.~A. Suarez, G.~W. Burr, M.~Mivelle, and
  D.~Charraut.
\newblock Full vectorial imaging of electromagneticlight at subwavelength
  scale.
\newblock {\em Opt. Express}, 18(6):5809--5824, 2010.

\bibitem{vo:ox12}
Thanh-Phong Vo, M~Mivelle, S~Callard, A~Rahmani, F~Baida, D~Charraut,
  A~Belarouci, D~Nedeljkovic, C~Seassal, GW~Burr, and T~Grosjean.
\newblock Near-field probing of slow bloch modes on photonic crystals with a
  nanoantenna.
\newblock {\em Opt. Express}, 20(4):4124--4135, 2012.

\bibitem{lindquist:scirep13}
Nathan~C Lindquist, Timothy~W Johnson, Prashant Nagpal, David~J Norris, and
  Sang-Hyun Oh.
\newblock Plasmonic nanofocusing with a metallic pyramid and an integrated
  c-shaped aperture.
\newblock {\em Scientific reports}, 3:1857, 2013.

\bibitem{xie:nl17}
Zhihua Xie, Yannick Lefier, Miguel~Angel Suarez, Mathieu Mivelle, Roland Salut,
  Jean-Marc Merolla, and Thierry Grosjean.
\newblock Doubly resonant photonic antenna for single infrared quantum dot
  imaging at telecommunication wavelengths.
\newblock {\em Nano Lett.}, 17(4):2152--2158, 2017.

\bibitem{kaschke:nanophotonics16}
Johannes Kaschke and Martin Wegener.
\newblock Optical and infrared helical metamaterials.
\newblock {\em Nanophotonics}, 5(4):510--523, 2016.

\bibitem{gibbs:apl13}
JG~Gibbs, AG~Mark, S~Eslami, and P~Fischer.
\newblock Plasmonic nanohelix metamaterials with tailorable giant circular
  dichroism.
\newblock {\em Appl. Phys. Lett.}, 103(21):213101, 2013.

\bibitem{esposito:acsphot14}
Marco Esposito, Vittorianna Tasco, Massimo Cuscuna?, Francesco Todisco, Alessio
  Benedetti, Iolena Tarantini, Milena~De Giorgi, Daniele Sanvitto, and Adriana
  Passaseo.
\newblock Nanoscale 3d chiral plasmonic helices with circular dichroism at
  visible frequencies.
\newblock {\em ACS Photon.}, 2(1):105--114, 2014.

\bibitem{esposito:natcom15}
Marco Esposito, Vittorianna Tasco, Francesco Todisco, Massimo Cuscun{\`a},
  Alessio Benedetti, Daniele Sanvitto, and Adriana Passaseo.
\newblock Triple-helical nanowires by tomographic rotatory growth for chiral
  photonics.
\newblock {\em Nat. Commun.}, 6:6484, 2015.

\bibitem{gansel:sci09}
Justyna~K Gansel, Michael Thiel, Michael~S Rill, Manuel Decker, Klaus Bade,
  Volker Saile, Georg von Freymann, Stefan Linden, and Martin Wegener.
\newblock Gold helix photonic metamaterial as broadband circular polarizer.
\newblock {\em Science}, 325(5947):1513--1515, 2009.

\bibitem{kosters:acsphot17}
Dolfine Kosters, Anouk De~Hoogh, Hans Zeijlemaker, Hakk{\i} Acar, Nir
  Rotenberg, and L~Kuipers.
\newblock Core--shell plasmonic nanohelices.
\newblock {\em ACS photon.}, 4(7):1858--1863, 2017.

\bibitem{schaferling:acsphot14}
Martin Sch{\"a}ferling, Xinghui Yin, Nader Engheta, and Harald Giessen.
\newblock Helical plasmonic nanostructures as prototypical chiral near-field
  sources.
\newblock {\em ACS Photon.}, 1(6):530--537, 2014.

\bibitem{wozniak:ox18}
Pawel Wozniak, Israel De~Leon, Katja Hoeflich, Caspar Haverkamp, Silke
  Christiansen, Gerd Leuchs, and Peter Banzer.
\newblock Chiroptical response of a single plasmonic nanohelix.
\newblock {\em Opt. Express}, 26(15):19275--19293, 2018.

\bibitem{wozniak:optica19}
Pawe{\l} Wo{\'z}niak, Israel De~Le{\'o}n, Katja H{\"o}flich, Gerd Leuchs, and
  Peter Banzer.
\newblock Interaction of light carrying orbital angular momentum with a chiral
  dipolar scatterer.
\newblock {\em Optica}, 6(8):961, 2019.

\bibitem{wang:lsa19}
Mengjia Wang, Roland Salut, Huihui Lu, Miguel-Angel Suarez, Nicolas Martin, and
  Thierry Grosjean.
\newblock Controlling light polarization by swirling surface plasmons.
\newblock {\em arXiv preprint arXiv:1812.06527}, 2018.

\bibitem{zhao:natnano17}
Yang Zhao, Amr~AE Saleh, Marie~Anne Van De~Haar, Brian Baum, Justin~A Briggs,
  Alice Lay, Olivia~A Reyes-Becerra, and Jennifer~A Dionne.
\newblock Nanoscopic control and quantification of enantioselective optical
  forces.
\newblock {\em Nat. Nanotechnol.}, 12(11):1055, 2017.

\bibitem{esposito:nl16}
Marco Esposito, Vittorianna Tasco, Francesco Todisco, Massimo Cuscunà, Alessio
  Benedetti, Mario Scuderi, Giuseppe Nicotra, and Adriana Passaseo.
\newblock Programmable extreme chirality in the visible by helix-shaped
  metamaterial platform.
\newblock {\em Nano Lett.}, 16(9):5823--5828, 2016.

\bibitem{balanis:book}
C.A. Balanis.
\newblock {\em Antenna theory: analysis and design}.
\newblock John Wiley \& Sons, New-York, 1997.

\bibitem{kraus:book}
John~D Kraus, Ronald~J Marhefka, and Ahmad~S Khan.
\newblock {\em Antennas and wave propagation}.
\newblock Tata McGraw-Hill Education, 2006.

\bibitem{banzer:ox10}
Peter Banzer, Jochen Kindler, Susanne Quabis, Ulf Peschel, and Gerd Leuchs.
\newblock Extraordinary transmission through a single coaxial aperture in a
  thin metal film.
\newblock {\em Opt. Express}, 18(10):10896--10904, 2010.

\bibitem{baida:apb07}
Fadi~I Baida.
\newblock Enhanced transmission through subwavelength metallic coaxial
  apertures by excitation of the TEM mode.
\newblock {\em Appl. Phys. B}, 89(2-3):145--149, 2007.

\end{thebibliography}

\end{document}